\documentclass[twocolumn,amsfonts,amsmath,amssymb,final,longbibliography,aps,prd,footinbib,nofootinbib,floatfix,10pt]{revtex4-1}
\usepackage[dvipsnames]{xcolor}
\usepackage{dcolumn}
\usepackage{lipsum}
\usepackage[colorlinks=true,allcolors=Blue,bookmarksopen=true,pdfencoding=auto]{hyperref}
\usepackage{upgreek,graphicx,manfnt,pifont,colonequals,bbm,stmaryrd}
\usepackage[shortlabels]{enumitem}
\usepackage{mathtools}
\usepackage{microtype}

\usepackage{amstext}
\usepackage{dsfont} 

\usepackage{titlesec}
\titleformat{\section}[runin]{\raggedright\bfseries}{\arabic{section}.}{1em}{}[.]
\titlespacing{\section}{0em}{1em}{0.5em}

\def\be{\begin{eqnarray}}
\def\ee{\end{eqnarray}}
\def\bea{\begin{eqnarray}}
\def\eea{\end{eqnarray}}

\DeclareMathOperator{\Res}{Res}
\DeclareMathOperator{\PE}{PE}
\newcommand{\ch}[1]{\chi_{\mathbf{#1}}}

\newcommand{\beal}{\begin{equation}
\begin{aligned}}
\newcommand{\eeal}{\end{aligned}
\end{equation}}
\newcommand{\bem}{\begin{multline}}
\newcommand{\eem}{\end{multline}}

\newcolumntype{L}[1]{>{\raggedright\let\newline\\\arraybackslash\hspace{0pt}}m{#1}}
\newcolumntype{C}[1]{>{\centering\let\newline\\\arraybackslash\hspace{0pt}}m{#1}}
\newcolumntype{R}[1]{>{\raggedleft\let\newline\\\arraybackslash\hspace{0pt}}m{#1}}

\begin{document}

\title{Superconformal indices for non-Lagrangian theories in five dimensions}

\medskip

\author{Hee-Cheol Kim${}^{a,b}$, Minsung Kim${}^{a}$, Sung-Soo Kim${}^{c}$ and Gabi Zafrir${}^{d}$}
 
\medskip

\medskip

\affiliation{${}^{a}$ Department of Physics, POSTECH, Pohang 790-784, Korea}
\affiliation{${}^{b}$ Asia Pacific Center for Theoretical Physics, Postech, Pohang 37673, Korea}
\affiliation{${}^{c}$ School of Physics, University of Electronic Science and Technology of China, No. 2006 Xiyuan Ave, West Hi-Tech Zone, Chengdu, Sichuan 611731, China}
\affiliation{${}^{d}$ Haifa Research Center for Theoretical Physics and Astrophysics, University of Haifa, Haifa 3498838, Israel}

 \date{\today}


\begin{abstract}
\noindent We propose two novel methods for computing the superconformal index of 5d superconformal field theories that cannot be described by conventional Lagrangian descriptions under mass deformations. The first approach involves the use of Higgs branch flows from UV Lagrangian theories, guided by transitions in 5-brane webs in Type IIB string theory. The second method employs the relationship between O$7^+$-plane and O$7^-$-plane with eight D7-branes, which applies to particular non-Lagrangian theories realized by brane configurations involving an O$7^+$-plane. As a concrete application of our method, we compute the superconformal indices for all known rank-1 non-Lagrangian theories, which we also use to identify flavor symmetries and their global forms at the conformal field theory (CFT) fixed points.
\end{abstract}

\maketitle

\subpdfbookmark{Introduction}{sec:intro}\section*{Introduction}
The superconformal index counts local BPS operators in a superconformal field theory (SCFT) \cite{Kinney:2005ej,Bhattacharya:2008zy}. As the index is a robust observable protected from any continuous deformations, it has been utilized to investigate non-perturbative phenomena, including dualities and symmetry enhancements, across various dimensions. In the case where the theory admits a weakly-coupled limit, the index can be computed from the partition function at the limit in radial quantization using supersymmetric localization techniques. There exist, however, numerous strongly-coupled isolated theories, such as the 5d rank-1 SCFT derived from M-theory compactified on a local $\mathbb{P}^2$ embedded in a Calabi-Yau (CY) threefold, that lack such weak coupling limits. For these theories, which we refer to as non-Lagrangian theories, the usual localization technique cannot be applied. In particular, at present there is no known way to compute superconformal indices for these non-Lagrangian theories in five dimensions.\footnote{In lower dimensions, various techniques have been developed to compute superconformal indices for non-Lagrangian theories. For instance, in the context of 4d theories, some notable examples can be found in \cite{Gadde:2015xta,Maruyoshi:2016tqk}.}

Non-Lagrangian theories have other BPS observables, for example, the partition function on $S^1$ times $\Omega$-deformed $\mathbb{R}^4$~\cite{Nekrasov:2002qd,Nekrasov:2003rj}, which can be computed on the Coulomb branch through expansions in terms of Coulomb branch moduli parameters. However, we cannot use these observables or the techniques used for them to calculate the superconformal index because the index requires integration over the Coulomb branch parameters. Also, while non-Lagrangian theories can be obtained by renormalization group (RG) flows from SCFTs that UV complete certain 5d gauge theories via large mass deformations, we cannot employ such RG flows to compute the IR index since large mass deformations are not feasible in the context of the superconformal index.

In this work, we present two approaches for evaluating the superconformal index of 5d non-Lagrangian theories. The first approach involves analyzing the RG flows originating from specific gauge theories on the Higgs branch and examining the corresponding limits of their superconformal indices.
In contrast to mass deformations, Higgs branch RG flows can be implemented at the level of superconformal index. At the end of Higgs branch RG flow from a UV theory, the index of the IR SCFT can be extracted from the residue of the UV index at a pole in the flavor fugacity associated to the Higgs branch deformation \cite{Gaiotto:2012xa}. 

Our first strategy to compute the superconformal index of a non-Lagrangian SCFT $\mathcal{T}_{IR}$ is to use the Higgs branch RG flows as follows. First, we will identify a gauge theory for a UV SCFT $\mathcal{T}_{UV}$ that can flow to $\mathcal{T}_{IR}$ via a Higgs branch RG flow, through a careful analysis of Type IIB $(p,q)$ 5-brane webs for the non-Lagrangian theory. Next, we will compute the superconformal index of $\mathcal{T}_{UV}$ using localization, and extract the residue at a specific pole in the index of $\mathcal{T}_{UV}$, which realizes the RG flow. Using this method, we will compute the superconformal indices of all rank-1 non-Lagrangian SCFTs, such as the local $\mathbb{P}^2$ theory, often referred to as the $E_0$ theory, and the local $\mathbb{P}^2+1{\bf Adj}$ theory \cite{Bhardwaj:2019jtr} which we denote as the $\widehat{E}_1$ theory. 

The second approach is to use the idea of ``{\it freezing}'' an O7$^-$-plane and eight D7-branes ($\text{O7}^- + 8\text{D7's}$) to an O7$^+$-plane in Type IIB 5-brane configurations, which was recently investigated in \cite{Hayashi:2023boy}. This idea can be implemented in the context of partition functions by carefully adjusting mass parameters or chemical potentials for the global symmetries associated with the D7-branes. This method offers a complementary method for computing the partition functions for the theories realized by brane configurations with an O7$^+$-plane. We extend this technique to calculate the superconformal indices of non-Lagrangian theories, including the $\widehat{E}_1$ theory. 

Our results will be verified by comparing the indices obtained from two separate techniques or distinct UV theories that yield equivalent non-Lagrangian SCFTs in the Higgs branch limits, and demonstrating that they agree. Additionally, we will use our findings to identify the correct flavor symmetries and their global forms at the UV fixed points.

\currentpdfbookmark{5-brane webs for non-Lagrangian theories}{sec:brane}\section*{5-brane webs for non-Lagrangian theories}
\begin{figure}[htbp]
  \centering
    \includegraphics[scale=0.27]{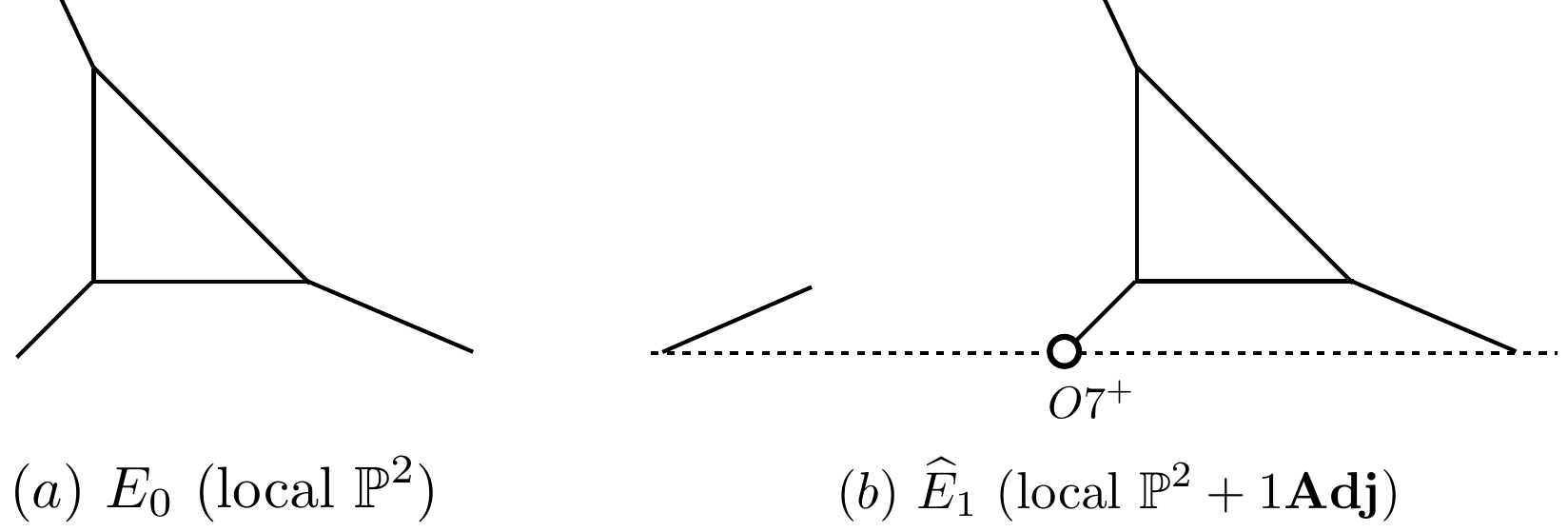}
    \caption{Rank-1 non-Lagrangian SCFTs. 
    An empty dot denotes the position of an O7$^+$ plane and the dotted line denotes its monodromy cut.
    }
    \label{F:rank-1-SCFTs}
\end{figure}

To begin, we 
consider 5-brane webs for 5d SCFTs on the Coulomb branch and explore how Higgsing transitions can lead to non-Lagrangian theories. At rank-1, which means one-dimensional Coulomb branch, there are two non-Lagrangian SCFTs whose 5-brane webs are depicted in FIG. \ref{F:rank-1-SCFTs}.
We first discuss the $E_0$ theory, which is the most basic rank-1 theory without flavor symmetry, corresponding to M-theory compactified on a local CY 3-fold containing a $\mathbb{P}^2$ surface, as
depicted in FIG. \ref{F:rank-1-SCFTs}(a). %
This theory is related to the SCFT that UV completes an $SU(2)$ gauge theory at the discrete theta $\theta=\pi$, through an RG-flow triggered by a mass deformation associated with its $U(1)$ symmetry. This deformation corresponds to the blow-down transition from a del Pezzo surface dP${}_1$ to $\mathbb{P}^2$ in geometry \cite{Morrison:1996xf,Douglas:1996xp}.

\begin{figure}[htbp]
  \centering
    \includegraphics[scale=0.22]{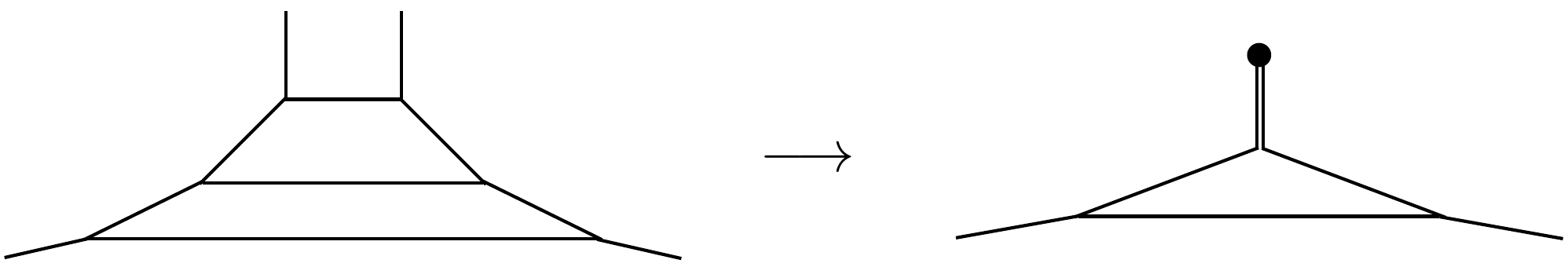}
    \caption{Higgs branch transition from the $SU(3)_3$ gauge theory (LEFT) to the $E_0$ theory (RIGHT) by attaching two parallel NS5-branes to a single 7-brane denoted by a solid dot.}
    \label{F:SU3-P2}
\end{figure}

We now illustrate how Higgs branch RG-flows from two distinct gauge theories lead to either the $E_0$ theory or two copies of it, which will later be utilized to calculate the superconformal index of the $E_0$ theory. Firstly, we can consider a Higgs branch limit of the $SU(3)_\kappa$ gauge theory at a Chern-Simons (CS) level $\kappa=3$. This theory has a $U(1)$ symmetry that is enhanced to an $SU(2)$ symmetry at the UV fixed point. By giving a vacuum expectation value (VEV) to a component of the moment map operator for the $SU(2)$ flavor symmetry, we can Higgs this theory to the $E_0$ theory. In the brane web, this Higgsing process can be achieved by anchoring two parallel NS5-branes on a single external 7-brane, as depicted in FIG. \ref{F:SU3-P2}. By moving the external 7-brane downward and performing a Hanany-Witten transition \cite{Hanany:1996ie}, we can observe that the resulting brane web is identical to the brane web for the $E_0$ theory.

\begin{figure}[htbp]
  \centering
    \includegraphics[scale=0.20]{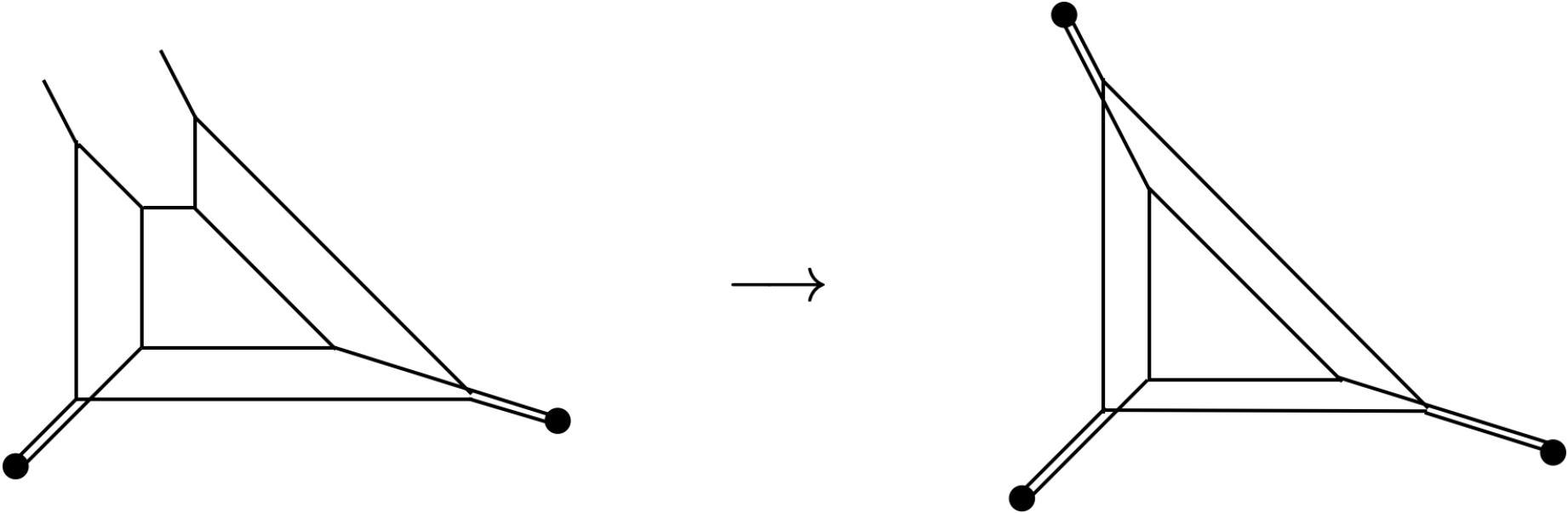}
    \caption{Higgs branch transition from the $SU(3)_6$ gauge theory (LEFT) to a pair of the $E_0$ theory (RIGHT) that are separated along the 7-branes. }
    \label{F:SU3-P2-2}
\end{figure}

The second UV theory that we will use for the index computation of $E_0$ is the $SU(3)_6$ gauge theory. The 5-brane web for this theory is illustrated on the left side of FIG. \ref{F:SU3-P2-2}. This theory also has a $U(1)$ symmetry which is enhanced to an $SU(2)$ flavor symmetry at the CFT fixed point. Again, we can Higgs this theory by giving a VEV to the moment map operator for the $SU(2)$ symmetry. In the IR limit, 
one obtains two copies of the $E_0$ theory which are decoupled from each other. This Higgs branch RG flow in the  brane web is demonstrated in FIG. \ref{F:SU3-P2-2}.

We now discuss the other rank-1 non-Lagrangian SCFT,  $\widehat{E}_1$ theory or the local $\mathbb{P}^2+1{\bf Adj}$ theory, which is derived from the 5d $\mathcal{N}=2$ $SU(2)$ gauge theory with $\theta=\pi$ by integrating out an instantonic hypermultiplet \cite{Bhardwaj:2019jtr} (see Appendix~\ref{sec:appendixC} for the precise relation). The corresponding 5-brane web  \cite{Kim:2020hhh} is given in FIG. \ref{F:rank-1-SCFTs}(b).

\begin{figure}[htbp]
  \centering
    \includegraphics[scale=0.187]{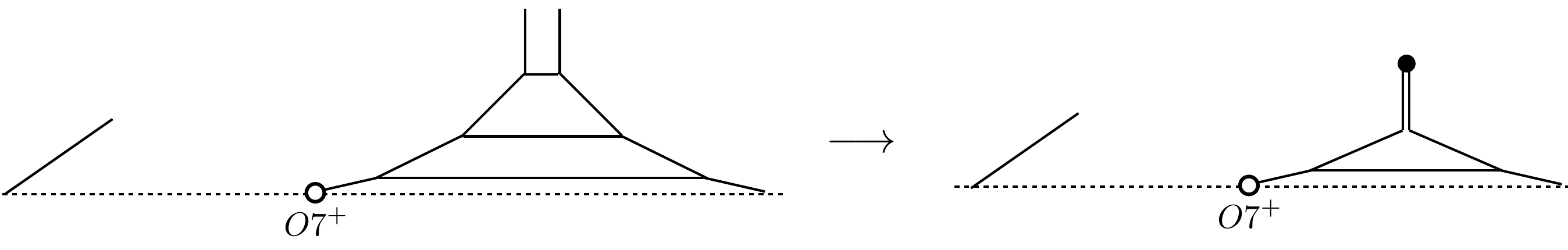}
    \caption{Transition from the $SU(3)_{\frac{1}{2}}+1{\bf Sym}$ to the $\widehat{E}_1$ theory.} \label{F:SU3-P2-adj}
\end{figure}

A UV theory that has a Higgs branch limit to the $\widehat{E}_1$ theory  is the SCFT UV completion of the $SU(3)_{\frac12}$ gauge theory with a symmetric hypermultiplet. The 5-brane web for this theory and its Higgsing process to the brane web for the $\widehat{E}_1$ theory are depicted in FIG. \ref{F:SU3-P2-adj}. As illustrated here, turning on a VEV for the moment map operator of the $SU(2)$ flavor symmetry and taking the low energy limit lead to the diagram on the right-hand side which is equivalent to the brane web for the $\widehat{E}_1$ theory through a Hanany-Witten transition \cite{Kim:2020hhh}.

In the following analysis, we will compute the superconformal indices for the $E_0$ theory and the $\widehat{E}_1$ theory using these UV gauge theories and examining their Higgs branch limits.

\currentpdfbookmark{Superconformal index and Higgsing}{sec:index}\section*{Superconformal index and Higgsing}
The 5d superconformal index is the Witten index of a 5d SCFT quantized on $S^4$. It is defined as \cite{Bhattacharya:2008zy,Kim:2012gu}
\begin{equation}
  \mathcal{I} = {\rm Tr}(-1)^F x^{2(j_r+R)}y^{2j_l} \textstyle\prod_i\mu_i^{F_i} \ ,  
\end{equation}
where $x,y$ are fugacities for the subgroup $SO(5)\times SU(2)_R$ of the superconformal group whose Cartan generators are $j_r,j_l,$ and $R$. $F_i$ denotes the flavor charge and $\mu_i$ is its fugacity.

The superconformal indices for 5d gauge theories with classical gauge groups can be computed by employing supersymmetric localization developed in \cite{Kim:2012gu}. This method, however, requires knowledge of the complete expression for the instanton partition function at every instanton order, which is essential for integrating over the gauge holonomies. This can be achieved through an ADHM construction for the instanton moduli space or by solving the blowup equations, as discussed in \cite{Nekrasov:2002qd,Nekrasov:2003rj,Nakajima:2003pg,Nakajima:2005fg}.  Unfortunately, this localization approach is not applicable to calculating superconformal indices for non-Lagrangian theories.  The primary reason for this is that the existing partition functions are expanded in terms of Coulomb branch parameters, which eventually need to be integrated to yield the superconformal index. Consequently, at present, there is no known method to calculate the superconformal index for any non-Lagrangian SCFT.

Now, we  introduce a simple method for calculating the superconformal index for non-Lagrangian SCFTs using the Higgs branch limits of certain UV gauge theories which we discussed in the previous section.  Specifically, we can take advantage of the fact that the instanton partition functions for the UV gauge theories we are studying can be explicitly computed using either the ADHM construction or the blowup approach. This allows us to compute their superconformal indices using localization. We then take the Higgs branch limit to the non-Lagrangian theory, which can be realized at the index level by taking the residue of the UV index at a specific pole, say $\xi=1$, associated to the moment map operator with a non-zero VEV \cite{Gaiotto:2012xa}. This procedure yields the superconformal index for the IR theory at the end of the RG flow, which can be written schematically as
\begin{align}
  \mathcal{I}[\mathcal{T}_{\rm IR}] = (\mathcal{I}_{\rm extra})^{-1} \times \Res_{\xi = 1} \mathcal{I}[\mathcal{T}_{\rm UV}] \ .
\end{align}
Here, $\mathcal{I}_{\rm extra}$ represents the contribution from the Goldstone modes of the broken flavor symmetry by the VEV. We multiplied
the inverse of $\mathcal{I}_{\rm extra}$ 
to subtract zero mode contributions under the Higgsing procedure.

As our first example, let us compute the superconformal index of the $E_0$ theory. To this end,  
we first calculate the index for the UV SCFT which leads to the $SU(3)_3$ gauge theory at low energy. Based on the localization argument, we can express the superconformal index for the $SU(3)_3$ gauge theory as an integral expression over the gauge holonomies $\alpha_i$ as
\begin{align}\label{eq:index-su3-3}
    &\mathcal{I}[SU(3)_3] = \frac{1}{3!}\! \oint \! \prod_{i=1}^2\frac{d\alpha_i}{2\pi}\, \mathcal{I}_{\rm pert}(\alpha_i;x,y) \!\cdot\! \mathcal{I}_{\rm inst}(\alpha_i,q;x,y) \ , \nonumber\\
    &\mathcal{I}_{\rm pert} \!=\! \prod_{e\in \Delta}\!\Big(2\sin\frac{e(\alpha)}{2}\Big)^2 \!\cdot\! \PE\!\left[\frac{-x(y\!+\!1/y)}{(1\!-\!xy)(1\!-\!x/y)}\!\sum_{e\in \Delta}\! e^{ie(\alpha)}\!\right]\!, \nonumber\\
    &\mathcal{I}_{\rm inst} = Z_{{\rm inst}}(\alpha_i,q;x,y)\cdot Z_{{\rm inst}}(-\alpha_i,q^{-1};x,y) \ ,
\end{align}
where PE[$f$] stands for the plethystic exponential of a single letter index $f$, $\Delta$ denotes the roots of the gauge group, and the contour encloses the unit circles $|e^{i\alpha_i}|=1$.
$\mathcal{I}_{\rm pert}$ is the perturbative contribution and the two $Z_{\rm inst}$ factors in $\mathcal{I}_{\rm inst}$ are the instanton partition functions capturing BPS spectrum of the $SU(3)_3$ theory on $\Omega$-deformed $S^1\times \mathbb{R}^4$ expanded by the instanton number fugacity $q$ and $q^{-1}$, respectively, which can be calculated using the standard ADHM construction for $SU(3)$ instantons. 

By collecting all the ingredients, we find the superconformal index for the 5d SCFT UV completion of the $SU(3)_3$ theory as
\begin{align}
    &\mathcal{I}[SU(3)_3] = 1 + \chi_{\bf 3}(q^{1/2})x^2 + \chi_{\bf 2}(y)\big[1 + \chi_{\bf 3}(q^{1/2})\big]x^3 \nonumber \\
    &\! + \big[\chi_{\bf 3}(y)\big(1+\chi_{\bf 3}(q^{1/2}))+1+\chi_{\bf 5}(q^{1/2})\big]x^4 \nonumber \\
    &\! +\big[ \chi_{\bf2}(y)\big(1\!+\!2\chi_{\bf3}(q^{1/2})+\chi_{\bf5}(q^{1/2})\big) \nonumber \\
    &\quad +\chi_{\bf4}(y)\big(1+\chi_{\bf3}(q^{1/2})\big)\big]x^5 +\mathcal{O}(x^6) \, ,
\end{align}
where $\chi_{\bf r}(a)$ stands for the ${\bf r}$-dimensional representation of the $SU(2)$ symmetry, e.g., 
$\chi_{\bf2}(a)\!=a+\!a^{-1}$. The term at $x^2$-order corresponds to the contribution from the moment map operators, which in turn reflects the enhancement of the flavor algebra to $\mathfrak{su}(2)$ at the CFT fixed point \cite{Bashkirov:2012re,Bergman:2013aca}. Also, we note that the index is expressed in the odd-dimensional characters for the flavor algebra. This is because the instanton partition function is expanded by $q$ and thus the index includes only terms with integral power of $q$. This implies that the flavor symmetry group is, in fact,  
$SU(2)/\mathbb{Z}_2=\!SO(3)$
rather than $SU(2)$ \cite{BenettiGenolini:2020doj,Apruzzi:2021vcu}.

The IR index for the $E_0$ theory can be obtained by tracing the Higgs branch limit of the index of the $SU(3)_3$ gauge theory. This limit corresponds to the moment map operator which develops a pole at $qx^2=1$. Therefore, to take the Higgs branch limit, we evaluate the residue of the UV index at $qx^2=1$.  By doing so, we obtain the superconformal index for the $E_0$ theory, 
\begin{align}\label{eq:index-p2}
    &\mathcal{I}[E_0] = (\mathcal{I}_{\rm extra})^{-1}\times \Res_{qx^2=1}\mathcal{I}[SU(3)_3] \nonumber \\
    &\quad = 1+ \chi_{\bf 2}(y)x^3 + \big[1+\chi_{\bf 3}(y)\big]x^4+\big[\chi_{\bf 2}(y)+\chi_{\bf 4}(y)\big]x^5 \nonumber \\
    &\quad +\big[-1+\chi_{\bf 3}(y)+\chi_{\bf 5}(y)\big]x^6+\big[\chi_{\bf 4}(y)+\chi_{\bf 6}(y)\big]x^7 \nonumber \\
    &\quad +\big[1+\chi_{\bf 3}(y)+2\chi_{\bf 5}(y)+\chi_{\bf 7}(y)\big]x^8 +\mathcal{O}(x^9) \ ,
\end{align}
where 
 \begin{align}\label{eq:goldstone}
   \mathcal{I}_{\rm extra} = \PE\left[ \frac{x^2}{(1-xy)(1-x/y)}\right] \ .
  \end{align}
Note that there is no term at $x^2$-order, which suggests that the IR theory lacks any conserved current. This result is indeed consistent with the absence of flavor symmetry in the $E_0$ theory.

The same result can be obtained by Higgsing from another UV theory, namely, the $SU(3)_6$ gauge theory. The superconformal index of the UV SCFT leading to the $SU(3)_6$ gauge theory under a mass deformation can also be expressed as the same form in (\ref{eq:index-su3-3}), but the instanton contributions are different. In this case, the instanton moduli space has no known ADHM construction\footnote{For the $SU(3)$ gauge theories with Chern-Simons level $\kappa > 3$, the $U(k)$ holonomy integral in the usual ADHM quantum mechanics of $k$ instantons exhibits higher order poles at asymptotic infinity. This indicates that the usual ADHM construction cannot provide consistent UV completions for the instantons in these theories.}, but we can instead compute the instanton partition function using the blowup equations as presented in \cite{Kim:2020hhh}. We compute the index of the 5d SCFT UV completion of the $SU(3)_6$ gauge theory as
\begin{align}
  &\mathcal{I}[SU(3)_6] = 1+ \chi_{\bf 3}(q)x^2 + \chi_{\bf 2}(y)\big(1+\chi_{\bf 3}(q)\big)x^3 \nonumber \\
  & \quad +\big[1+\chi_{\bf 2}(y)\chi_{\bf 2}(q)+\chi_{\bf 3}(y)(1+\chi_{\bf 3}(q))+\chi_{\bf 5}(q)\big]x^4 \nonumber \\
  & \quad +\big[\chi_{\bf2}(y)(1+2\chi_{\bf3}(q)+\chi_{\bf 5}(q))+\chi_{\bf2}(q)(1+\chi_{\bf3}(y))\big]x^5 \nonumber \\
  & \quad+\chi_{\bf4}(y)\big(1+\chi_{\bf3}(q)\big)x^5 \!+\!\mathcal{O}(x^6)\ .
\end{align}
The result shows that the flavor symmetry of the $SU(3)_6$ gauge theory at the CFT fixed point is $SU(2)$. 

By evaluating the residue of this index at $q^2x^2=1$, one can compute the superconformal index of the IR theory, which consists of two decoupled $E_0$ theories, as we previously illustrated in FIG. \ref{F:SU3-P2-2} using 5-brane webs. Thus, we arrive at the following relation:
\begin{align}
  \left(\mathcal{I}[E_0]\right)^2 &= (\mathcal{I}_{\rm extra})^{-1}\times \Res_{q^2x^2= 1} \mathcal{I}[SU(3)_6] \ .
\end{align}
We verified that this relation agrees well with the index of the $E_0$ theory given in (\ref{eq:index-p2}) up to the $x^9$-order, which provides a strong evidence for our result.

Next, we shall compute the superconformal index of the $\widehat{E}_1$ theory using the Higgsing procedure described in FIG. \ref{F:SU3-P2-adj}. To begin, we first compute the index of the $SU(3)_{\frac{1}{2}}+1{\bf Sym}$ theory at the UV fixed point. The superconformal index of this UV theory can be written as
\begin{align}\label{eq:index-su3-sym}
    &\mathcal{I}\big[SU(3)_{\frac12}\!+\!1{\bf Sym}\big] \!=\! \frac{1}{3!}\! \oint \! \prod_{i=1}^2\frac{d\alpha_i}{2\pi}\, \mathcal{I}_{\rm pert}\cdot \mathcal{I}_{\rm inst} \ ,\\
    &\mathcal{I}_{\rm pert} \!=\! \prod_{e\in \Delta}\!\Big(2\sin\frac{e(\alpha)}{2}\Big)^2\cdot \PE \bigg[\frac{-x(y+1/y)}{(1\!-\!xy)(1\!-\!x/y)}\sum_{e\in \Delta}\! e^{ie(\alpha)} \nonumber\\
    &\quad  \quad +\!\frac{x}{(1\!-\!xy)(1\!-\!x/y)}\!\sum_{w\in {\bf Sym}}\!(e^{iw(\alpha)+i\mu}+e^{-iw(\alpha)-i\mu})\bigg]\! ,\nonumber \\
    &\mathcal{I}_{\rm inst} = Z_{{\rm inst}}(\alpha_i,\mu,q;x,y)\cdot Z_{{\rm inst}}(-\alpha_i,-\mu,q^{-1};x,y) \ , \nonumber
\end{align}
where $\mu$ denotes the chemical potential for the $U(1)$ flavor symmetry of the symmetric hypermultiplet. Here, the instanton partition function can be calculated using the ADHM construction for $SU(3)$ instantons coupled to additional matter in the form of a symmetric hypermultiplet, which we summarize in Appendix~\ref{sec:appendixA}. By evaluating the integral, we find
\begin{align}\label{eq:index-su3-sym2}
  &\mathcal{I}\!\big[SU(3)_{\frac12}\!+\!1{\bf Sym}\big] \!=\! 1+\big[\chi_{\bf3}(a)+\chi_{\bf3}(b)\big]x^2 \\
  &\ \ +\big[\chi_{\bf2}(y)(1+\chi_{\bf3}(a)+\chi_{\bf3}(b))+\chi_{\bf 4}(a)\chi_{\bf3}(b)\big]x^3 \nonumber \\
  & \ \ +\big[\chi_{\bf3}(y)(1+\chi_{\bf3}(a)+\chi_{\bf3}(b))+\chi_{\bf2}(y)\chi_{\bf4}(a)\chi_{\bf3}(b)\big]x^4 \nonumber \\
  & \ \ +\big[2\ch{5}(a) + \ch{5}(b) + \ch{3}(a) \ch{3}(b) + 3 \big]x^4 +\mathcal{O}(x^5) ,\nonumber 
\end{align}
where $a\equiv qe^{i\mu/2}$ and $b \equiv \sqrt{q}e^{-5i\mu/4}$. This result enables us to identify the precise global flavor symmetry group of the UV CFT as $SU(2)_a\times SO(3)_b$, which also confirms the $\mathfrak{so}(4)$ flavor algebra in this theory predicted in \cite{Bhardwaj:2020avz}.

We now take the Higgs branch limit by giving a VEV to the moment map operator for the $SO(3)_b$ flavor symmetry. To achieve this at the level of the index, we compute the residue of the UV index at the pole $b^2x^2=1$. The result then leads to the superconformal index for the $\widehat{E}_1$ theory as follows:
\begin{align}\label{eq:index-E1hat}
  \mathcal{I}[\widehat{E}_1] 
   & \!=\! 1 \!+\! \widehat{\chi}_{\bf3}\, x^2 \!+\! \big[\chi_{\bf2}(y)(1\!+\!\widehat{\chi}_{\bf3})\!+\!\widehat\chi_{\bf4}\big]x^3 \!+\!\mathcal{O}(x^4) ,   
\end{align}
where $\widehat{\chi}_{\bf r}\!\equiv\!{\chi}_{\bf r}(a)$ with $a$ being the fugacity for the flavor symmetry.
The coefficient $\widehat{\chi}_{\bf 3}=a^2\!+\!1\!+\!a^{-2}$ 
at $x^2$-order captures the contribution of the conserved current multiplet. From this, we can conclude that the naive $U(1)_a$ flavor symmetry group is enhanced to $SU(2)_a$  at the CFT fixed point.

We can try to test \eqref{eq:index-E1hat} by comparing the result against a different construction of the 5d SCFT. One such realization is through the dimensional reduction of the 6d $A_2$ $(2,0)$ theory with an outer automorphism twist \cite{Bhardwaj:2019jtr}. While we cannot compute the index using this realization, we can understand some general features of its BPS spectrum. In particular, we can argue that the basic BPS multiplets in this theory should be the $SU(2)$ conserved current multiplets and a Higgs branch operator in the ${\bf 4}$ of the $SU(2)$ flavor symmetry and whose ground state is in the ${\bf 4}$ of $SU(2)_R$. This indeed matches what we observe in \eqref{eq:index-E1hat}. See Appendix~\ref{sec:appendixC} for a more detailed analysis.   

We have computed the above index for $\widehat{E}_1$ up to $x^3$-order using the instanton partition function of the UV $SU(3)_{\frac{1}{2}}+1{\bf Sym}$ theory up to 3-instantons. However, to precisely determine higher order terms of the $x$-expansion, we need higher instanton results, which our computational resources unfortunately cannot currently accommodate. In what follows, as an alternative approach, 
we will introduce another technique to compute the superconformal index of specific theories engineered by brane webs with an O7$^+$-plane. By employing this method, we will improve the order of the index computation for the $\widehat{E}_1$ theory.


\currentpdfbookmark{The freezing branes and \^{E}1 theory}{sec:new}\section*{The freezing branes and $\widehat{E}_1$ theory} 
For theories involving an O7$^+$-plane in their brane configuration, such as $SU(N)$ theory with symmetric matter or $SO(N)$ theory, an alternative yet novel approach for computing physical observables was proposed in \cite{Hayashi:2023boy}. 
This method suggests that the O7$^+$-plane contribution can be retrieved from ``{\it freezing}'' an O7$^-$-plane and eight D7-branes. It is a particular tuning of the hypermultiplet masses with some special values such that most of the contributions from the eight fundamental hypermultiplets (D7-branes) cancel each other and what remains, combined with the contribution of an O7$^-$-plane,  yields the contribution of an O7$^+$-plane. For instance, the $SU(N)+1\mathbf{Sym}$ theory can be effectively treated as the $SU(N)+1\mathbf{AS}+8\mathbf{F}$ theory by such freezing procedure. This was explicitly checked at the level of the Seiberg-Witten curves \cite{Hayashi:2023boy} and can be extended to the instanton partition functions \cite{KLNY:2023}. It should be noted that not all physical observables associated with an O7$^+$-plane can be obtained from freezing O7$^-+8$D7s. Only for some observables like the prepotential or index function, it would provides an effective way of computing them. See \cite{KLNY:2023} for more details on its limitation.

It is also worth noting that the freezing is applicable to non-Lagrangian theories. An illustrative example discussed in \cite{Hayashi:2023boy} is the $\widehat{E}_1$ theory, which can be seen as a local $\mathbb{P}^2+1\mathbf{AS}+8\mathbf{F}$ theory with appropriately adjusted mass parameters. Since the antisymmetric hypermultiplet decouples here, the $\widehat{E}_1$ theory can be effectively obtained by freezing the $Sp(1)+7\mathbf{F}$ theory.

We accordingly implement this freezing to compute the index for the $\widehat{E}_1$ theory. Recall that the superconformal index of the UV SCFT of $ Sp(1)+7\mathbf{F} $ is given by a gauge holonomy integral expression as \cite{Kim:2012gu, Hwang:2014uwa},
\begin{align}
    &\mathcal{I}[Sp(1)\!+\!7\mathbf{F}] = \frac{1}{2!} \oint \frac{d\alpha}{2\pi} \mathcal{I}_{\mathrm{pert}} \cdot \mathcal{I}_{\mathrm{inst}} \, , \label{Sp1+7F-index-integral}\\
    &\mathcal{I}_{\mathrm{pert}} = \Big(2\sin\frac{\alpha}{2}\Big)^2 \cdot \PE\bigg[\frac{-x(y\!+\!1/y)}{(1\!-\!xy)(1\!-\!x/y)} \sum_{e \in \Delta} e^{ie(\alpha)} \nonumber \\
    &\ + \frac{x}{(1\!-\!xy)(1\!-\!x/y)} \sum_{l=1}^7 \sum_{w \in \mathbf{F}} \big(e^{iw(\alpha)+i\mu_l} \!+\! e^{-iw(\alpha)-i\mu_l}\big) \bigg] , \nonumber\\
    &\mathcal{I}_{\mathrm{inst}} = Z_{\mathrm{inst}}(\alpha,\mu_l,q;x,y) \cdot Z_{\mathrm{inst}}(-\alpha,-\mu_l,q^{-1},x,y) \, , \nonumber
\end{align}
where $ \mu_l $ are the chemical potentials for fundamental hypers, and $ Z_{\mathrm{inst}} $ is the instanton partition function which can be computed using the ADHM construction \cite{Kim:2012gu, Hwang:2014uwa}. We have carried out the explicit integration and computed this index up to $x^{10}$-order. The detailed result can be found in Appendix \ref{sec:appendixFreezing}.

Next, we proceed with the freezing procedure acting on this index, which eventually gives rise to the index of the $\widehat{E}_1$ theory. To accomplish this, we adjust the flavor fugacities, following the process discussed in \cite{Hayashi:2023boy, KLNY:2023}, as
\begin{alignat}{2}\label{eq:tuning}
    \begin{aligned}
        &e^{i\mu_{1,2}^*} \mapsto a x^{\pm 1/2} \, , \quad
        &&e^{i\mu_{3,4}^*} \mapsto a y^{\pm 1/2} \, , \\
        &e^{i\mu_{5,6}^*} \mapsto -a x^{\pm 1/2} \, , \quad
        &&e^{i\mu_{7,8}^*} \mapsto -a^{\pm1} y^{1/2} \, ,
    \end{aligned}
\end{alignat}
where $ a$ is the fugacity for the flavor symmetry of the  $ \widehat{E}_1 $ theory. We then obtain with this tuning the superconformal index of the $ \widehat{E}_1 $ theory,
\begin{align}\label{eq:freezingE1hat}
    &\mathcal{I}[Sp(1)\!+\!7\mathbf{F}] \overset{\mu_l\rightarrow \mu_l^*}{\mapsto} \PE\bigg[\frac{-x(a^2+a^{-2})}{(1-xy)(1-x/y)}\bigg] \cdot \mathcal{I}[\widehat{E}_1] \, , \nonumber  \\
    &\mathcal{I}[\widehat{E}_1] \!=\! 1 \!+\! \widehat{\chi}_{\bf 3} x^2 \!+\! \Big[\ch{2}(y) (1+\widehat{\chi}_{\bf 3}) \!+\! \widehat{\chi}_{\bf 4}\Big]x^3 \nonumber \\
    &\ +\! \Big[\ch{3}(y)(1+\widehat{\chi}_{\bf 3}) \!+\! \ch{2}(y)\big(\widehat{\chi}_{\bf 2}\!+\!\widehat{\chi}_{\bf 4}\big) 
    +\!\widehat{\chi}_{\bf 5}\!+\!2    \Big]x^4 
    \nonumber \\
    &\ +\! \Big[\ch{4}(y)(1\!+\!\widehat{\chi}_{\bf 3})\! +\! \ch{3}(y)(\widehat{\chi}_{\bf 2}\!+\!\widehat{\chi}_{\bf 4}) \!+\! \ch{2}(y)(2\!+\! 2\widehat{\chi}_{\bf 3} \!+\! \widehat{\chi}_{\bf 5})\nonumber \\
    &\quad +\! \widehat{\chi}_{\bf 2} + \widehat{\chi}_{\bf 4} + \widehat{\chi}_{\bf 6}\Big] x^5 + \mathcal{O}(x^6) \, .
\end{align}
This perfectly agrees with the result in \eqref{eq:index-E1hat} and further improves it up to $x^5$-order.


\currentpdfbookmark{Summary}{sec:summary}\section*{Summary}
We have introduced novel method to compute the superconformal indices of 5d non-Lagrangian SCFTs. Our methods involve using Higgs branch RG flows from a UV Lagrangian theory or employing a freezing procedure on orientifold 7-brane in brane constructions. By applying these techniques, we are able to calculate the superconformal indices for rank-1 non-Lagrangian SCFTs, namely the $E_0$ theory and the $\widehat{E}_1$ theory, given in 
\eqref{eq:index-p2} and \eqref{eq:freezingE1hat}, respectively.


\section*{Acknowledgments}
We are grateful to Hirotaka Hayashi, Kimyeong Lee, Xiaobin Li, Satoshi Nawata, Jaewon Song, and Futoshi Yagi for their insightful discussions and comments. SK thanks the hospitality of POSTECH where part of this work was done. HK and MK are supported by Samsung Science and Technology Foundation under Project Number SSTF-BA2002-05 and by the National Research Foundation of Korea (NRF) grant funded by the Korea government (MSIT) (2023R1A2C1006542). SK is supported by the NSFC grant No. 12250610188.

\vspace{0.5cm}







\appendix

\titleformat{\section}[runin]{\raggedright\bfseries}{Appendix \Alph{section}: }{0em}{}[.]
\section{Comparison of the \texorpdfstring{$\widehat{E}_1$}{\^{E}1} index with a 6d realization} \label{sec:appendixC}

We can try to test our result for the index of the $\widehat{E}_1$ theory by comparing against expectations from other constructions. One such construction, is the realization of the $\widehat{E}_1$ theory from the twisted compactification of the $A_2$ $6d$ $(2,0)$ theory on a circle. Indeed, this 
was how the $\widehat{E}_1$ theory was originally found in \cite{Bhardwaj:2019jtr}.

To illustrate how the $\widehat{E}_1$ theory arises in such a construction, it is convenient to start first with a simpler example of the reduction of the $A_1$ 6d $(2,0)$ theory on a circle. In this case, we expect to get, at low-energies, the 5d maximally supersymmetric $SU(2)$ gauge theory with discrete theta angle $\theta=0$ \cite{Douglas:2010iu,Lambert:2010iw,Tachikawa:2011ch} (we shall henceforth use the shorthand notation $SU(2)_{\theta}$ for the $SU(2)$ gauge theory with theta angle $\theta$). We can think of this theory also as a minimally supersymmetric 5d $SU(2)$ gauge theory with an adjoint hypermultiplet. In this viewpoint, we have an $SU(2)$ flavor symmetry, rotating the hyper, which is the commutant of the $SU(2)$ R-symmetry of 5d minimal SUSY in the $USp(4)$ R-symmetry of 5d maximal SUSY. When performing the circle reduction, we have the freedom of incorporating holonomies in flavor symmetries on the circle. Of particular interest is the holonomy inside this $SU(2)$ subgroup of the $USp(4)$ R-symmetry group, which is viewed as a flavor symmetry in minimal SUSY. This will lead to a 5d theory with minimal SUSY. 

When the holonomy is small, it simply leads to the adjoint hyper acquiring a mass. This leads to the 5d version of the $\mathcal{N}=2^*$ theory, which in the deep IR reduce to a pure minimally supersymmetric 5d $SU(2)_0$ gauge theory. Of course, this holds when the holonomy is small, but what happens when the holonomy becomes large? To better understand this, it is convenient to take a closer look at the BPS operators in the theory. Our starting point is the 6d $A_1$ $(2,0)$ theory. The main BPS multiplet in this theory is the $D_1 [0,0,0]^{(0,2)}_4$, where we use the notation of \cite{Cordova:2016emh}, which is the $(2,0)$ energy-momentum tensor multiplet. Its ground state is a scalar in the ${\bf 14}$ of $USp(4)$. We would be interested in what happens to this state when we dimensionally reduce. For this, it is convenient to first decompose $USp(4)\rightarrow SU(2)_R \times SU(2)_F$, with $SU(2)_R$ the R-symmetry of the minimally supersymmetric case. Under this decomposition we have that: ${\bf 14}_{USp(4)} \rightarrow 1 + {\bf 2}_{SU(2)_R} {\bf 2}_{SU(2)_F} + {\bf 3}_{SU(2)_R} {\bf 3}_{SU(2)_F}$. Here we shall only consider the last term, which corresponds to the shortest BPS multiplet we get in 5d from this 6d multiplet\footnote{This follows from the shortening conditions, where acting on the ground state with the supercharge annihilates the state where the Dynkin indices of the $USp(4)$ representations of the supercharge and the ground state are summed \cite{Cordova:2016emh}. In other words, trying to raise the weight of the $USp(4)$ representation of the ground state by acting with the supercharge annihilates the state. If we now look at the highest weight representation under $SU(2)_R$ in the decomposition, then it will obey a similar shortening condition as trying to raise its $SU(2)_R$ state is tantamount to raising its $USp(4)$ highest weight.}. This operator is simply the conserved current multiplet of the 5d $SU(2)_F$ symmetry.

When we reduce on a circle, we get the 5d operator, as well as all its Kaluza-Klein excitations. An interesting aspect in 5d gauge theories resulting from the compactification of 6d SCFTs is that they generally retain some knowledge on the KK tower through instanton particles, see for instance \cite{Douglas:2010iu,Lambert:2010iw,Tachikawa:2015mha}. As such we would want to keep track on the entire KK tower. For this purpose, we introduce two fugacities, $f$ and $q$, with the first one associated with the holonomy for $SU(2)_F$ and the second to the radius of the circle. Note then, that the first keep track of the mass to the adjoint hyper, while the second keeps track of the mass of the KK modes. In the 5d gauge theory, these are associated with background central charges to the $SU(2)_F$ symmetry and the $U(1)$ instanton symmetry of the gauge theory, respectively, justifying their designation as fugacities. All in all, we can write the charges of the expected states in 5d as:
\begin{align}
\Big(1\!+\!f\!+\!\frac{1}{f}\Big)\!\!\! \sum^{\infty}_{n=-\infty}\!\! q^n \!\rightarrow\! \Big(1\!+\!f\!+\!\frac{1}{f}\Big) \! \Big(1\!+\!q\!+\!\frac{1}{q}\!+\!q^2\!+\!\frac{1}{q^2}\!+\!\cdots\! \Big)
\end{align}
Note that all these states carry the charges of conserved currents in 5d. Next consider the case where we reduce on a circle of finite radius but without a holonomy. In that case, all states charged under $q$ acquire a mass, and the currents associated with them are no longer conserved. We then end up with the $SU(2)_F$ conserved current of the 5d maximally supersymmetric gauge theory. 

If we next introduce a small holonomy, then we break $SU(2)_F$ to $U(1)$, and correspondingly the currents charged under $f$ will no longer be conserved. This describes the regime where we get in 5d the minimally supersymmetric $SU(2)_0$ gauge theory. However, if we continue to increase the holonomy, we will eventually reach a state where $m_f = m_q$ and the states with charges $\frac{f}{q}$ become massless. At this point we expect to get a new 5d theory due to the presence of additional massless degrees of freedom. Said new theory should have an $SU(2)$ global symmetry, due to the extra currents, be minimally supersymmetric and support a mass deformation leading to the 5d gauge theory. Indeed there is a theory that ticks all these boxes: the $E_1$ SCFT. Therefore, it is natural to expect that at this regime of the compactification data we get the $E_1$ 5d SCFT.

\begin{figure}[ht]
\center
\includegraphics[width=0.48\textwidth]{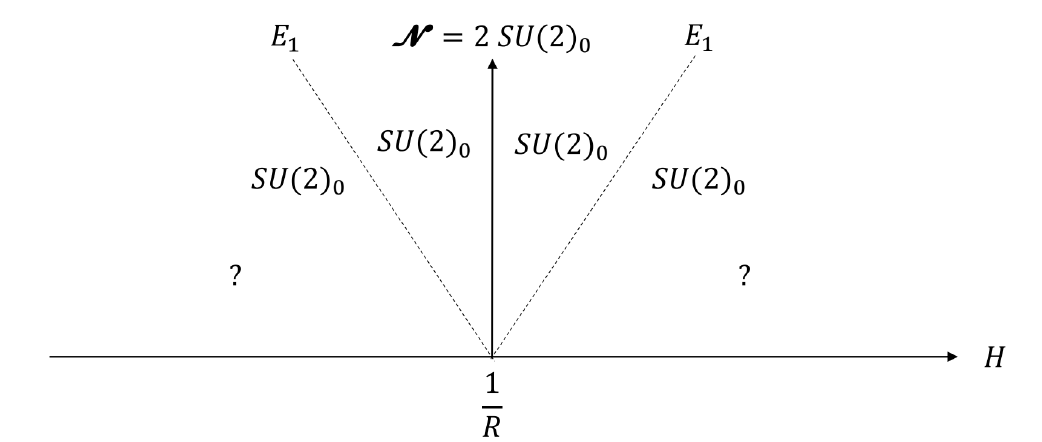} 
\caption{A schematic summery of the expected low-energy theory for the compactification of the 6d $(2,0)$ $A_1$ theory. The vertical axis stands for the inverse of the compactification radius, which sets the scale for the KK masses. The horizontal axis stands for the mass deformation associated with the holonomy in the $USp(4)$ R-symmetry. We use $\mathcal{N}=2$ $SU(2)_{\theta}$ for the maximally 5d supersymmetric $SU(2)$ gauge theory with the listed $\theta$ angle, and $SU(2)_{\theta}$ for the minimally supersymmetric one. Here, we do not explore the full parameter space, rather contenting ourselves with some region around the maximal SUSY case. Domains lying outside of said region, where the behavior remains unexplored, are symbolized with a "?".}
\label{E1R}
\end{figure}

What happens if we increase $m_f$ further? The $E_1$ SCFT is known to support only a single mass deformation sending it to the 5d $SU(2)_0$ gauge theory for both positive and negative mass. As such we should again get the gauge theory, at least as long as $m_f$ does not increase too much. The expected picture of the resulting low-energy theory in these cases is shown in figure \ref{E1R}. We could continue exploring other possible phase transition though that would not be of interest to us here. We also note that there could be in principle phase transitions due to massless matter coming from longer multiplets, though again we shall not pursue this here.

Now we are going to turn to the case of actual interest to us, the circle reduction of the $A_2$ $(2,0)$ theory. This theory is known to have a $\mathbb{Z}_2$ discrete symmetry associated with the outer automorphism of $A_2$, and we shall be mainly interested in the case where the reduction is done with a twist in said symmetry. In this case we get, at low-energies, the 5d maximally supersymmetric $SU(2)_{\pi}$ gauge theory \cite{Tachikawa:2011ch}. This is again if we compactify without an $SU(2)_F$ holonomy, so maximal SUSY is preserved. Next we would be interested in the result when we also include this holonomy. Again when the holonomy is small, it will simply imply a mass term for the adjoint hyper and we expect to get the 5d minimally supersymmetric $SU(2)_{\pi}$ gauge theory. We would want to know what happens when the holonomy becomes large.

To understand this, we again turn to look at the BPS multiplets in the 6d SCFT. Now we have two of them. One is still the energy-momentum tensor multiplet, in the $D_1 [0,0,0]^{(0,2)}_4$ of the $(2,0)$ superconformal group, while the other is in the $D_1 [0,0,0]^{(0,3)}_6$ \footnote{For a generic $(2,0)$ theory of type $G$, there would be $rank(G)$ such operators in the $D_1 [0,0,0]^{(0,d_i)}_{2d_i}$, for $d_i$ the dimensions of the invariant polynomials of $G$.}. We shall refer to them as $V_2$ and $V_3$, respectively. Their ground state are again scalars, with the $V_2$ one being in the ${\bf 14}$ of $USp(4)$ as before, while the $V_3$ one being in the ${\bf 30}$ of $USp(4)$. When decomposed into the $SU(2)_R \times SU(2)_F$ subgroup, we have that: ${\bf 30}_{USp(4)} \rightarrow 1 + {\bf 2}_{SU(2)_R} {\bf 2}_{SU(2)_F} + {\bf 3}_{SU(2)_R} {\bf 3}_{SU(2)_F} + {\bf 4}_{SU(2)_R} {\bf 4}_{SU(2)_F}$, where again we would only be interested in the last state, which carries the highest weight under $SU(2)_R$.

When reduced, $V_2$ would give us the same states as before, including the KK tower, so we turn now to $V_3$. One interesting property of this multiplet is that it is odd under the $\mathbb{Z}_2$ symmetry we twist by. As such it essentially obeys anti-periodic boundary conditions on the circle, and its KK masses are fractional compared with those of the periodic fields. If we write in terms of fugacities we have:
\begin{align}
    & \Big(1\!+\!f\!+\!\frac{1}{f}\Big) \!\!\! \sum^{\infty}_{n=-\infty} \!\! q^n \!\rightarrow\! \Big(1\!+\!f\!+\!\frac{1}{f}\Big)\! \Big(1\!+\!q\!+\!\frac{1}{q}\!+\!q^2\!+\!\frac{1}{q^2}\!+\!\cdots\!\Big) \nonumber \\
    & \Big(f^{\frac{1}{2}}\!+\!\frac{1}{f^{\frac{1}{2}}} \!+\! f^{\frac{3}{2}} \!+\! \frac{1}{f^{\frac{3}{2}}}\Big) \!\!\sum^{\infty}_{n=-\infty} \!\! q^{n+\frac{1}{2}} \\
    & \quad \rightarrow\! \Big(f^{\frac{1}{2}}\!+\!\frac{1}{f^{\frac{1}{2}}} \!+\! f^{\frac{3}{2}}\!+\!\frac{1}{f^{\frac{3}{2}}}\Big) \Big(q^{\frac{1}{2}}\!+\!\frac{1}{q^{\frac{1}{2}}} \!+\! q^{\frac{3}{2}}\!+\!\frac{1}{q^{\frac{3}{2}}}\!+\!\cdots\!\Big) \, , \nonumber
\end{align}
where the first term are the states due to $V_2$, while the second are from $V_3$. In the latter, we have the charges in the ${\bf 4}$ of $SU(2)_F$ which multiplies the KK tower, which now carries fractional powers in $q$ due to the twist. We also note that the first term has 5d superconformal charges of a conserved current, while the other has those of a Higgs branch operator whose lowest component is a scalar in the ${\bf 4}$ of $SU(2)_R$.

Next we inquire what happens as we increase the value of the holonomy. As before, we expect to get the 5d $SU(2)_{\pi}$ gauge theory as long as the holonomy isn't big enough that additional states in the KK tower become massless. This occurs first when $3 m_f = m_q$, in which case the states with the charges $\frac{q^{\frac{1}{2}}}{f^{\frac{3}{2}}}$ becomes massless. We note, though, that no additional state in the conserved current multiplet becomes massless. As such, the global symmetry should remain $U(1)$. We again expect that in this regime we get a new minimal SUSY theory with a $U(1)$ global symmetry and that possess a mass deformation leading to the 5d gauge theory. Like in the previous case, such a theory indeed exists, which is the so-called the $\tilde{E}_1$ SCFT. This theory appears as the UV completion of the 5d $SU(2)_{\pi}$ gauge theory.

What happens if we continue to increase the holonomy? As in the previous case, we now trigger the same mass deformation but in the opposite direction. However, unlike the previous case, the low-energy theory one gets for the $\tilde{E}_1$ SCFT depends on the sign of the mass deformation, with one sign giving the 5d gauge theory while the other giving the $E_0$ SCFT. Since we got to this theory from the gauge theory side, the other direction must lead to the $E_0$ theory. Therefore, we conclude that once the holonomy increases we should get the $E_0$ SCFT at low-energies. 

What happens if we continue to increase the holonomy? The next point where we get massless matter is when $m_f = m_q$. However, now the massless matter we get are the $\frac{q}{f}$ ones from the current multiplet, but also $\frac{q^{\frac{1}{2}}}{f^{\frac{1}{2}}}$ and $\frac{q^{\frac{3}{2}}}{f^{\frac{3}{2}}}$ from $V_3$. As we now have extra conserved currents, we should get an $SU(2)$ global symmetry, and so expect to get a new 5d theory with $SU(2)$ global symmetry and a deformation leading to the $E_0$ 5d SCFT. This is the $\widehat{E}_1$ 5d SCFT. We also notice that we get additional massless states from $V_3$. These should provide a Higgs branch operator in the ${\bf 4}$ of the $SU(2)$, and whose ground state is a scalar in the ${\bf 4}$ of $SU(2)_R$. Such a state should contribute to the index as $\chi_{\bf 4}(c)x^3$. This is precisely what we observe in \eqref{eq:index-E1hat}, with this state and the $SU(2)$ conserved currents being the first low-lying states we observe. The fact that we can identify the origin of these operators, using a different realization of the $\widehat{E}_1$ theory, is an indication in favor of the correctness of \eqref{eq:index-E1hat}.

\begin{figure}[h]
\center
\includegraphics[width=0.48\textwidth]{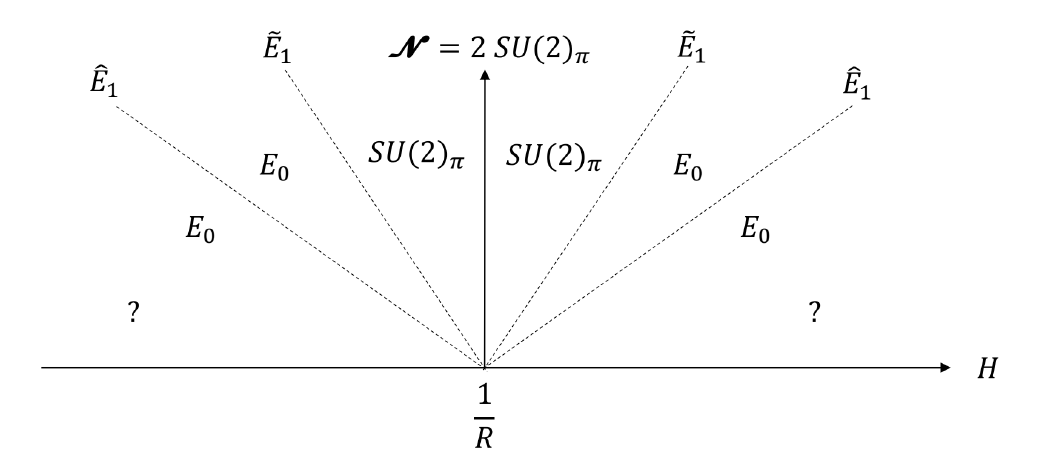} 
\caption{A schematic summery of the expected low-energy theory for the compactification of the 6d $(2,0)$ $A_2$ theory with a twist in its outer automorphism symmetry. Here we use the same notations as in Figure \ref{E1R}.}
\label{E1tR}
\end{figure}

To complete the discussion, we can again ask what happens if we further increase the holonomy? Again, this will trigger the mass deformation of the $\widehat{E}_1$ SCFT. Like in the $E_1$ case, this mass deformation is in an $SU(2)$ global symmetry and so the Weyl group of $SU(2)$ relates the positive and negative mass deformations. As such increasing the holonomy should again yield the $E_0$ SCFT as the low-energy theory. Figure \ref{E1tR} provides a summary of the different theories we get in this case. We can in principle consider increasing the mass deformation further, though we shall not pursue this here.



\section{Partition functions of \texorpdfstring{$SU(3)_{\frac{1}{2}}+1{\bf Sym}$}{SU(3)1/2+1sym}} \label{sec:appendixA}
We summarize the detailed computations for the partition functions of $ SU(3)_{\frac{1}{2}} + 1\mathbf{Sym} $ theory. We first compute the instanton partition function on $\Omega$-deformed $ S^1\times \mathbb{C}^2$ using the ADHM construction for the instanton moduli space which generalizes the discussions in \cite{Nekrasov:2002qd,Nekrasov:2003rj,Hwang:2014uwa}. The instanton partition function is expressed as a series $ Z_{\mathrm{inst}}=\sum_k q^k Z_k $ expanded by the instanton number fugacity $ q $, and $ Z_k $ is the $ k $-instanton partition function. In the case of $ SU(N)_\kappa $ gauge theories, $ Z_k $ takes a contour integral expression as
\begin{align}\label{ADHM-integral}
    Z_k = \frac{1}{k!} \oint \left[\prod_{I=1}^k \frac{du_I}{2\pi i}\right] e^{-\kappa \sum_{I=1}^k u_I} Z_{\mathrm{vec}} \prod_{\mathbf{R}} Z_{\mathbf{R}} \, ,
\end{align}
where $ \kappa $ is the Chern-Simons level, $ Z_{\mathrm{vec}} $ and $ Z_{\mathbf{R}} $ are the contributions from the vector multiplet and the hypermultiplets in representation $ \mathbf{R} $, respectively. $ Z_{\mathrm{vec}} $ is given by
\begin{align}
    Z_{\mathrm{vec}} &= \frac{\prod_{I,J} 2\sinh(\frac{u_{IJ}^-+2\epsilon_+}{2}) \cdot \prod_{I \neq J} 2\sinh(\frac{u_{IJ}^-}{2})}{\prod_{I,J} 2\sinh(\frac{u_{IJ}^- + \epsilon_{1,2}}{2}) \cdot \prod_{I,j} 2\sinh(\frac{\pm (u_I - a_j) + \epsilon_+}{2})} \, ,
\end{align}
and the contribution from a symmetric hypermultiplet is given by
\begin{align}
    Z_{\mathbf{Sym}} \!\!=\!\! \frac{\prod_{I,j}\! 2\sinh(\frac{u_I + a_j + m_1}{2})\! \prod_{I \leq J}\! 2\sinh(\frac{\pm(u_{IJ}^{+}+m_1)-\epsilon_{\!-\!}}{2})\!}{\prod_{I<J}\! 2\sinh(\frac{\pm(u_{IJ}^++m_1)-\epsilon_+}{2})}.
\end{align}
Here, $ u_{IJ}^\pm=u_I\pm u_J $, $\epsilon_{1,2}$ are $\Omega$-deformation parameters with  $\epsilon_\pm = \frac{\epsilon_1\pm\epsilon_2}{2} $, $ a_j $ are $ SU(N) $ gauge holonomies satisfying $ \sum_{j=1}^N a_j = 0 $, and $ m_1 $ is the mass parameter of the symmetric hyper. The contour integral in \eqref{ADHM-integral} is evaluated through the Jeffrey-Kirwan (JK) residue prescription \cite{Hwang:2014uwa}.

However, the integral in \eqref{ADHM-integral} for $ SU(3)_{\frac{1}{2}}\!+\!1\mathbf{Sym} $ exhibits poles at the infinity $ u_I = \pm \infty $ with degree higher than 1. Treating these higher degree poles poses a subtle issue, and currently, we lack a suitable method to handle such poles in contour integrals. To avoid this subtlety, we introduce an additional antisymmetric hypermultiplet ($\bf AS$) and first compute the instanton partition function of the $ SU(3)_0+1\mathbf{Sym}+1\mathbf{AS}$ theory. The contour integral receives an additional contribution from the antisymmetric hyper with mass $m_2$ as
\begin{equation}
    Z_{\mathbf{AS}} \!\!=\!\! \frac{\prod_{\substack{I,j}} 2\sinh(\frac{u_I + a_j + m_2}{2})\! \prod_{I<J}\! 2\sinh(\frac{\pm(u_{IJ}^++m_2)-\epsilon_-}{2})\!}{\prod_{I\leq J} 2\sinh(\frac{\pm(u_{IJ}^++m_2)-\epsilon_+}{2})} \ .
\end{equation}
This theory with $1{\bf AS}$ is a 5d KK theory and the theory $ SU(3)_{\frac{1}{2}}+1\mathbf{Sym} $ which we are interested in can be obtained by integrating out the antisymmetric matter.

One can evaluate the contour integral of  the $ SU(3)_0+1\mathbf{Sym}+1\mathbf{AS}$ theory using the JK-residue prescription, and we checked that the resulting instanton partition function becomes a solution to the blowup equation introduced in \cite{Kim:2020hhh} up to 3-instanton order. To compute the instanton partition function of the $ SU(3)_{\frac{1}{2}}+1\mathbf{Sym} $ theory, we take the limit $ m_2 \to \infty $. Then we find
\begin{align}
    & \lim_{m_2\rightarrow \infty} Z_{\rm inst}[SU(3)_0 + 1\mathbf{Sym} + 1\mathbf{AS}] \\
    &\to \PE\left[\frac{-x^2 e^{-5i\mu/2}}{(1-xy)(1-x/y)}\right] \cdot Z_{{\rm inst}}[SU(3)_{\frac{1}{2}} + 1\mathbf{Sym}] \nonumber
\end{align}
and
\begin{align}\label{sym-xexpand}
    &Z_{{\rm inst}}[SU(3)_{\frac{1}{2}} + 1\mathbf{Sym}] = \PE\left[\sum_{k=1}^\infty q^k \mathcal{Z}_k\right] \,, 
\end{align}
with the first few $\mathcal{Z}_k$ being 
\begin{widetext}
\begin{align}
    &\mathcal{Z}_1 \!=\! \big(e^{-5i\mu/2} \!+\! \chi_{\overline{\mathbf{3}}}^{A_2} e^{3i\mu/2} \big)x^2 \!+\! \Big[\ch{2}(y) \big(e^{-5i\mu/2} \!-\! \chi_{\mathbf{3}}^{A_2} e^{-i\mu/2} \!+\! \chi_{\overline{\mathbf{3}}}^{A_2} e^{3i\mu/2}\big) \!+\! e^{i\mu/2} \!-\! \chi_{\mathbf{3}}^{A_2} e^{5i\mu/2}\Big]x^3 \\
    &\ +\! \Big[\ch{3}(y) \big(e^{-5i\mu/2} \!-\! \ch{\mathbf{3}}^{A_2}e^{-i\mu/2} \!+\! \chi_{\overline{\mathbf{3}}}^{A_2}e^{3i\mu/2} \big) \!+\! \ch{2}(y) \big((\ch{8}^{A_2}\!+\!2)e^{i\mu/2} \!-\! \ch{3}^{A_2}e^{5i\mu/2} \!+\! \chi_{\overline{\mathbf{3}}}^{A_2}e^{-3i\mu/2} \big) \!-\! \big(\chi_{\overline{\mathbf{6}}}^{A_2}\!+\!3\chi_{\mathbf{3}}^{A_2}\big) e^{-i\mu/2}\Big] x^4 \nonumber \\
    &\ +\! \Big[\ch{4}(y)\big( e^{-5i\mu/2} \!-\! \chi_{\mathbf{3}}^{A_2} e^{-i\mu/2} \!+\! \chi_{\overline{\mathbf{3}}}^{A_2} e^{3i\mu/2} \big)\!+\! \ch{3}(y) \big( (\chi_{\mathbf{8}}^{A_2}\!+\!2) e^{i\mu/2} \!-\! \chi_{\mathbf{3}}^{A_2} e^{5i\mu/2} \!+\! \chi_{\overline{\mathbf{3}}}^{A_2}e^{-3i\mu/2} \big) 
    \nonumber \\
    &\quad~ 
    \!-\! \ch{2}(y) \big( ( \chi_{\mathbf{15}}^{A_2} \!+\! 2\chi_{\overline{\mathbf{6}}}^{A_2} \!+\! 4\chi_{\mathbf{3}}^{A_2}) e^{-i\mu/2} \!+\! \chi_{\mathbf{6}}^{A_2} e^{3i\mu/2} \big) \!+\! \big(\chi_{\mathbf{10}}^{A_2} \!+\! 3\chi_{\mathbf{8}}^{A_2}\!+\!1\big)e^{i\mu/2} \!+\! \big(\chi_{\overline{\mathbf{15}}}^{A_2} \!+\! 2\chi_{\overline{\mathbf{3}}}^{A_2}\big)e^{-3i\mu/2}\Big] x^5 + \mathcal{O}(x^6) \, , \nonumber \\
    &\mathcal{Z}_2 \!=\! e^{i\mu}x^2 \!+\! \left(\ch{2}(y) e^{i\mu} \!+\! e^{-2i\mu} \!+\! e^{4i\mu}\right) x^3 \!+\! \Big[ \ch{3}(y) e^{i\mu} \!+\! \big(\ch{2}(y) \big( e^{4i\mu} \!+\! e^{-2i\mu} \!-\! \chi_{\overline{\mathbf{3}}}^{A_2} e^{2i\mu} \big) \!+\! e^{i\mu} \!-\! \chi_{\overline{\mathbf{3}}}^{A_2} e^{-i\mu} \big) \Big] x^4  \nonumber \\
    &\ +\! \Big[ \ch{4}(y) e^{i\mu} \!+\! \ch{3}(y) \big( e^{4i\mu} \!+\! e^{-2i\mu} \!-\! \chi_{\overline{\mathbf{3}}}^{A_2} e^{2i\mu} \big) \!+\!\ch{2}(y) \big( (\chi_{\mathbf{8}}^{A_2}\!+\!3)e^{i\mu} \!+\! \chi_{\mathbf{3}}^{A_2} e^{3i\mu} \!-\! \chi_{\overline{\mathbf{3}}}^{A_2} e^{-i\mu}\big) \!-\! \big(\chi_{\mathbf{6}}^{A_2} \!+\! 3\chi_{\overline{\mathbf{3}}}^{A_2}\big)e^{2i\mu} \Big] x^5 \!+\! \mathcal{O}(x^6) \, , \nonumber \\
    &\mathcal{Z}_3 \!=\! e^{3i\mu/2} x^3 \!+\! \ch{2}(y) e^{3i\mu/2} x^4 \!+\! \Big[ \ch{3}(y) e^{3i\mu/2} \!-\! \big(\chi_{\overline{\mathbf{3}}}^{A_2} e^{-i\mu/2} \!+\! e^{3i\mu/2}\big) \Big] x^5 \!+\! \mathcal{O}(x^6) \, , \nonumber
\end{align}
\end{widetext}
where $ x=e^{-\epsilon_+} $, $y=e^{-\epsilon_-} $, $ \mu=im_1 $, and $ \chi_{\mathbf{R}}^{A_2} $ is the character of representation $ \mathbf{R} $ in $ \mathfrak{su}(3) $ gauge algebra.

Next, we compute the superconformal index using the expression in \eqref{eq:index-su3-sym}.
However, our current computational resources impose limitations on our ability to compute the instanton partition function only up to 3-instantons. This result may not be sufficient for us to obtain the superconformal index beyond the third order in $x$-expansion. To improve our computational capacity, we first identify the flavor symmetry of the SCFT, which turns out to be $SU(2)_a\times SO(3)_b$, using the result up to $x^2$-order. Then we exploit the fact that the index must form representations of the flavor symmetry. This approach enables us to compute the higher-order terms up to $x^6$-order as follows:
\begin{align}
    &\mathcal{I}[SU(3)_{\frac{1}{2}}\!+\!1\mathbf{Sym}]
    \!=\! 1 \!+\! (\ch{3}(a) \!+\! \ch{3}(b)) x^2  \\
    &\!+\! (1\!\!+\!\!\ch{3}(a)\!\!+\!\!\ch{3}(b)) (\ch{2}(y) x^3 \!\!+\!\! \ch{3}(y) x^4 \!\!+\!\! \ch{4}(y) x^5 \!\!+\!\! \ch{5}(y) x^6) \nonumber \\
    &\!+\! \ch{4}(a) \ch{3}(b) (x^3 \!+\! \ch{2}(y) x^4 \!+\! \ch{3}(y) x^5 \!+\! \ch{4}(y) x^6) \nonumber \\
    &\!+\! (2\ch{5}(a) \!+\! \ch{5}(b) \!+\! \ch{3}(a) \ch{3}(b) \!+\! 3) x^4 \!+\! 3\ch{2}(y)x^5 \nonumber \\
    &\!+\! \ch{2}(y) [2\ch{5}(a) \!+\! \ch{5}(b) \!+\! (2\ch{3}(a)\!+\!1)\ch{3}(b) \!+\! 3\ch{3}(a)] x^5 \nonumber \\
    &\!+\! [(\ch{6}(a)\!+\!2\ch{4}(a)\!+\!\ch{2}(a)) \ch{3}(b) \!+\! \ch{4}(a) \ch{5}(b) \!-\! \ch{2}(a) ]x^5 \nonumber \\
    &\!+\! \ch{3}(y) [3\ch{5}(a)\!+\!2\ch{5}(b)\!+\!(\ch{3}(a)\!+\!1)(3\ch{3}(b)\!+\!4)\!+\!2] x^6 \nonumber \\
    &\!+\! \ch{2}(y) [(2\ch{6}(a)\!+\!5\ch{4}(a)\!+\!3\ch{2}(a)) \ch{3}(b) \!-\! \ch{2}(a)]x^6 \nonumber \\
    &\!+\! [\ch{2}(y)\ch{4}(a)(2\ch{5}(b)\!+\!1) \!+\! (2\ch{5}(a)\!+\!\ch{3}(a)\!+\!5)\ch{3}(b) ] x^6 \nonumber \\
    & \!+\! [(\ch{7}(a)\!+\!2\ch{3}(a))(\ch{5}(b)\!+\!3) \!+\! \ch{7}(b) \!+\! 2\ch{5}(a)] x^6 \!\!+\!\! \mathcal{O}(x^7) . \nonumber
\end{align}
Obviously, the higher instanton contributions to this result involves some guess work relying on representations of the $SU(2)_a\times SO(3)_b$ flavor symmetry. Nevertheless, we will confirm this result by Higgsing to the index of the $\widehat{E}_1$ theory, and comparing it with another result obtained by a freezing procedure, which we will carry out below.

We compute the superconformal index of the $ \widehat{E}_1 $ theory by taking Higgs branch limit corresponding to extracting residue at the pole $bx^2\!=\!1$ of the above index. The result is
\begin{align}
    \mathcal{I}[\widehat{E}_1] =\frac{  \Res_{bx^2=1} \mathcal{I}[SU(3)_{\frac{1}{2}}+1\mathbf{Sym}]}{\mathcal{I}_{\mathrm{extra}}\cdot \mathcal{I}_{\mathrm{extra}}'} ,
\end{align}
where
\begin{align}
    \mathcal{I}_{\mathrm{extra}}' &\!=\! \PE\bigg[\frac{x(a^3 \!+\! a \!+\! a^{-1} \!+\! a^{-3})}{(1\!-\!xy)(1\!-\!x/y)}\bigg]
\end{align}
corresponds to contributions from free hypermultiplets that decouple from the CFT.

\section{\texorpdfstring{$\widehat{E}_1$}{\^{E}1} index from freezing} \label{sec:appendixFreezing}
We now discuss the freezing procedure to calculate the superconformal index of the $ \widehat{E}_1 $ theory from the index of the $Sp(1)+7\mathbf{F}$ theory. 
The superconformal index of  $Sp(1)+7\mathbf{F} $ was previously computed in \cite{Kim:2012gu, Hwang:2014uwa}. We further improve those results  and compute higher order terms as follows:

\begin{widetext}
    \begin{align}\label{7F-index}
        &\mathcal{I}[Sp(1)\!+\!7\mathbf{F}] = 1 \!+\! \chi_{\mathbf{248}}^{E_8} x^2 \!+\! (\ch{248}^{E_8}\!+\!1) (\ch{2}(y) x^3 \!+\! \ch{3}(y)x^4 \!+\! \ch{4}(y) x^5 \!+\! \cdots \!+\! \ch{9}(y)x^{10}) \!+\! (\ch{27000}^{E_8}\!+\!1) x^4 \\
    & \! +\! \ch{2}(y)(\ch{30380}^{E_8} \!+\! \ch{27000}^{E_8} \!+\! \ch{248}^{E_8} \!+\! 1)x^5  \!+\! (\ch{3}(y) (\ch{30380}^{E_8} \!+\! 2\ch{27000}^{E_8} \!+\! \ch{3875}^{E_8} \!+\! 2\ch{248}^{E_8} \!+\! 2) \!+\! \ch{1763125}^{E_8} \!+\! \ch{30380}^{E_8} \!+\! 2\ch{248}) x^6 \nonumber \\
    & \!+\! (\ch{4}(y)(2\ch{30380}^{E_8} \!+\! 2\ch{27000}^{E_8} \!+\! \ch{3875}^{E_8} \!+\! 4\ch{248}^{E_8} \!+\! 2) \!+\! \ch{2}(y)(\ch{4096000}^{E_8} \!+\! \ch{1763125}^{E_8} \!+\! \ch{30380}^{E_8} \!+\! 2\ch{27000}^{E_8} \!+\! \ch{3875}^{E_8} \!+\! 4\ch{248} \!+\! 2) )  x^7 \nonumber \\
    &\! +\! (\ch{5}(y) (2\ch{30380}^{E_8} \!+\! 3\ch{27000}^{E_8} \!+\! 2\ch{3875}^{E_8} \!+\! 5\ch{248}^{E_8} \!+\! 4) \! +\! \ch{3}(y) (2\ch{4096000}^{E_8} \!+\! 2\ch{1763125}^{E_8} \!+\! \ch{779247}^{E_8} \!+\! 3\ch{30380}^{E_8} \!+\! 3\ch{27000}^{E_8}))x^8 \nonumber \\
    & \!+\! ( \ch{3}(y) (\ch{3875}^{E_8} \!+\! 8\ch{248}^{E_8} + 3) \!+\! \ch{79143000}^{E_8} \!+\! \ch{4096000}^{E_8} \!+\! \ch{2450240}^{E_8} \!+\! \ch{30380}^{E_8} \!+\! 3\ch{27000}^{E_8} \!+\! \ch{3875}^{E_8} \!+\! 2\ch{248}^{E_8} \!+\! 3)x^8  \nonumber \\
    &\! +\! (\ch{6}(y)(3\ch{30380}^{E_8} \!+\! 3\ch{27000}^{E_8} \!+\! 2\ch{3875}^{E_8} \!+\! 7\ch{248}^{E_8}\!+\!4) \!+\! \ch{4}(y)(3\ch{4096000}^{E_8} \!+\! \ch{2450240}^{E_8} \!+\! 3\ch{1763125}^{E_8} \!+\! 2\ch{779247}^{E_8} \!+\! 5\ch{30380}^{E_8}))x^9 \nonumber \\
    &\!+\! (\ch{4}(y)(7\ch{27000}^{E_8} \!+\! 3\ch{3875}^{E_8} \!+\! 11\ch{248}^{E_8} \!+\! 7) \!+\! \ch{2}(y)(\ch{281545875}^{E_8} \!+\! \ch{79143000}^{E_8} \!+\! 3\ch{4096000}^{E_8} \!+\! \ch{2450240}^{E_8} \!+\! 2\ch{1763125}^{E_8}))x^9 \nonumber \\
    &\!+\! \ch{2}(y)(2\ch{779247}^{E_8} \!+\! \ch{147250}^{E_8} \!+\! 5\ch{30380}^{E_8} \!+\! 6\ch{27000}^{E_8} \!+\! 2\ch{3875}^{E_8} \!+\! 8\ch{248}^{E_8} \!+\! 4) x^9 \!+\! \ch{7}(y)(3\ch{30380}^{E_8} \!+\! 4\ch{27000}^{E_8} \!+\! 3\ch{3875}^{E_8} )x^{10} \nonumber \\
    &\!+\! (\ch{7}(y)(8\ch{248}^{E_8} \!+\! 6) \!+\! \ch{5}(y)(5\ch{4096000}^{E_8} \!+\! \ch{2450240}^{E_8} \!+\! 4\ch{1763125}^{E_8} \!+\! 4\ch{779247}^{E_8} \!+\! \ch{147250}^{E_8} \!+\! 9\ch{30380}^{E_8} \!+\! 10\ch{27000}^{E_8} \!+\! 4\ch{3875}^{E_8}))x^{10} \nonumber \\
    &\!+\! (\ch{5}(y) (18\ch{248}^{E_8} \!+\! 8) \!+\! \ch{3}(y) (2\ch{281545875}^{E_8} \!+\! \ch{203205000}^{E_8} \!+\! 2\ch{79143000}^{E_8} \!+\! \ch{70680000}^{E_8} \!+\! 6\ch{4096000}^{E_8} \!+\! 3\ch{2450240}^{E_8})) x^{10} \nonumber \\
    &\!+\! (\ch{3}(y)(4\ch{1763125}^{E_8} \!+\! 4\ch{779247}^{E_8} \!+\! 2\ch{147250}^{E_8} \!+\! 11\ch{30380}^{E_8} \!+\! 15\ch{27000}^{E_8} \!+\! 7\ch{3875}^{E_8} \!+\! 15\ch{248}^{E_8} \!+\! 10) \!+\! \ch{2642777280}^{E_8} \!+\! 9\ch{248}^{E_8} \!+\! 2) x^{10} \nonumber \\
    &\!+\! (\ch{344452500}^{E_8} \!+\! \ch{281545875}^{E_8} \!+\! 3\ch{4096000}^{E_8} \!+\! 4\ch{1763125}^{E_8} \!+\! 3\ch{779247}^{E_8} \!+\! \ch{147250}^{E_8} \!+\! 6\ch{30380}^{E_8} \!+\! 4\ch{27000}^{E_8} \!+\! 2\ch{2875}^{E_8}) x^{10}  \!+\! \mathcal{O}(x^{11}) \, , \nonumber
    \end{align}
\end{widetext}
where $ \chi_{\mathbf{r}}^{E_8} = \chi_{\mathbf{r}}^{E_8}(e^{i\mu_1},\cdots,e^{i\mu_8}) $ are the characters of representation $ \mathbf{r}$ of the $E_8$ flavor symmetry at the CFT fixed point, as predicted in \cite{Seiberg:1996bd}, and $ e^{i\mu_8} \equiv q^{-2} $ is the instanton number fugacity. 
To arrive at this result, we used the instanton partition function $ Z_{\mathrm{inst}} $ up to 5-instantons. With this, we can determine the index only up to $ x^7$-order. 
To compute the terms beyond $ x^7$-order, even though it requires higher instanton contributions, we use the fact that the terms in the index must form $ E_8 $ representations. This allows us to uniquely fix all higher-order terms up to $x^{10}$-order. 

We will now proceed with the freezing procedure, which effectively converts the brane configuration with an O7$^-$-plane with 8 D7-branes for the $Sp(1)+7\mathbf{F}$ theory into that with an O7$^+$-plane and eventually leads to the brane web for $\widehat{E}_1$ as investigated in \cite{Hayashi:2023boy, KLNY:2023}.  In the context of the index, this corresponds to the specialization of $E_8$ fugacities in the index of $Sp(1)+7\mathbf{F}$, as given by \eqref{eq:tuning}. Consequently, we can compute the index for the $\widehat{E}_1$ theory from that for the $Sp(1)+7\mathbf{F}$ by applying the specialization of $E_8$ fugacities.
The index of $Sp(1)+7\mathbf{F}$ up to $x^{10}$-order in \eqref{7F-index} enables us to calculate the superconformal index of the $\widehat{E}_1$ theory up to $x^5$-order which is summarized in \eqref{eq:index-E1hat}. To obtain the result beyond this order, we need higher instanton computations at $k>5$ of the $Sp(1)+7\mathbf{F}$ theory, which unfortunately are not currently available.

\bibliography{refs}
\end{document}